\documentclass[12pt,thmsa]{article}
\usepackage{graphicx}

\begin{document}

\author{D. B. Ion and M. L. D. Ion}
\title{\textbf{Principle of minimum complexity as a new principle in hadronic
scattering}}
\date{2 Dec 2005 }
\maketitle

\begin{abstract}
In this paper a measure of complexity of the system of [$J$ and $\theta $%
]-quantum states produced in hadronic scattering is introduced in terms of
the $[S_{J}(p)$, $S_{\theta }(q)]-$scattering entropies. We presented strong
experimental evidence for the \textit{saturation of the} $%
[S_{J}^{o1}(p),S_{\theta }^{o1}(q)]-$\textit{PMD-SQS optimal limits in }the
pion-nucleon and kaon-nucleon scatterings.\textit{\ }The validity of \textit{%
a principle of minimum complexity in hadronic interaction} is
supported with high accuracy ($CL>99\%$) by the experimental data
on pion-nucleon especially at energies higher than 2 GeV.
\end{abstract}

\section{\textbf{Introduction}}

In this paper, by using the optimality [1], the scattering entropies [2-4] $%
[S_{J}(p)$, $S_{\theta }(q)],$ \textit{\ }a measure of the complexity [5] of
quantum scattering in terms of Tsallis-like entropies [6] is proposed. Then,
the \textit{nonextensive statistical behavior, optimality [1] and complexity
of the }[$J$\textit{\ and }$\theta $]\textit{-quantum states} in hadronic
scatterings are investigated in an unified manner. A connection between
optimal states obtained from the \textit{principle of minimum distance in
the space of quantum states} (PMD-SQS)[1] and the most stringent (MaxEnt)
entropic bounds on Tsallis-like entropies for quantum scattering is
established. A measure of the complexity of quantum scattering in terms of
Tsallis-like entropies is proposed. The results on the experimental tests of
\textit{the PMD-SQS-optimality}, as well as on the\textit{\ complexity},
obtained by using the experimental pion-nucleon pion-nucleus phase shifts%
\textit{, }are presented. Then, the nonextensivity indices p and q are
determined from the experimental entropies by a fit with the optimal
entropies $[S_{J}^{o1}(p)$, $S_{\theta }^{o1}(q)]$ obtained from the\textit{%
\ principle of minimum distance in the space of states}. In this
way strong experimental evidences for the $p-$nonextensivities in
the range $0.5\leq p\leq 0.6$ with $q$ correlated via relation
$1/p+1/q=2,$ are obtained with high accuracy ($CL>99\%$) from the
experimental data of the pion-nucleon and pion-nucleus
scatterings.

\section{\textbf{What is the complexity of a quantum system of scattering
states? }}

In the last time many authors have developed and commented on the
multiplicity and variety of definitions of \textit{complexity (C)} in the
scientific literature (see Ref.[5] and references therein).

What is the \textit{complexity} of an interacting system? In
general complexity is a concept very difficult to be defined. If
we consider that this notion is coming from the original Latin
word \textit{complexus}, which signifies ''\textit{entwined}'',
''\textit{twisted together}'' then this may interpreted in the
following way: in order to have a complex system you need two or
more distinct components, which are joined in such a way that it
is difficult to separate them. If we look in the Oxford Dictionary
we find that \textit{something is defined as complex if it is
''made of closely connected parts''. }So, in these definitions we
find the basic duality between parts which are at the same time
distinct and connected. This implies that: a system would be more
complex if more parts could be distinguished, and if more
connections between them exist. So, the aspects of
\textit{distinction} and \textit{connections} between the parts of
systems determine two dimensions characterizing complexity.
Distinction corresponds to the fact that different parts of the
complex behaves differently while connections correspond to
constrains. Moreover, the \textit{distinction} leads at the
limits to \textit{disorder} $(\Delta ),$ or chaos, while the \textit{%
connection} leads to order $(\Omega ).$ Complexity can only exist if both
aspects are present: neither perfect disorder nor perfect order are complex.
Before of a precise mathematical definition of the \textit{complexity
measure (C) }it is useful to point out some intuitive\textit{\ }properties
of the \textit{complexity} of the physical systems [5]. These are as follows:

(a) Any reasonable\textit{\ measure of complexity must vanish for the
complete ordered} or \textit{complete disordered states of the system; }

(b) Any \textit{\ complexity must be an universal physical
quantity}$-$that is, it is applied to any dynamical system;

(c) Any \textit{complexity must correspond to mathematical complexity in
problem solving}.

Current definitions may be divided into three distinct categories according
to the following essential feature:\ (a) Definitions which take \textit{%
complexity} as \textit{monotonically increasing function} of \textit{%
disorder (}$\Delta $); (b) Complexity defined as a \textit{convex function}
of \textit{disorder} \textit{(}$\Delta $); Complexity defined as \textit{%
monotonically increasing function} of \textit{order} \textit{(}$\Omega
=1-\Delta $).

Since we will express our \textit{complexity measure (C) }in terms
of \textit{disorder (}$\Delta $), or of \textit{order} ($\Omega
$), we start with the definitions [5]
\begin{equation}
\Delta =\frac{S}{S^{\max }},\smallskip \ \Omega =1-\Delta =1-\frac{S}{%
S^{\max }}  \label{1}
\end{equation}
where S is the Boltzmann-Gibs-Shannon (BGS) entropy
\begin{equation}
S=-\sum p_{i}\ln p_{i}  \label{2}
\end{equation}
where p$_{i},1\leq i\leq N,$ is the probability of i-state of the
N states available to the system (the Boltzmann constant k is
taken k=1). In the simplest case of an isolated system, the
maximum entropy  occurs to the equiprobable distribution
$p_{i}=1/N,$ $1\leq i\leq N,$ yielding
\begin{equation}
S_{\max }=\ln N  \label{3}
\end{equation}

All categories of \textit{complexity measures} can be included by
defining a measure of the form [5]
\begin{equation}
C^{\alpha \beta }(q)=\Delta ^{\alpha }\Omega ^{\beta }  \label{4}
\end{equation}
So, the \textit{complexity measures }of category I and III correspond to the
particular cases: $\alpha >0$, $\beta =0,$and $\alpha =0,\beta >0.$ The
category II is obtained when both $\alpha >0$ and $\beta >0.$ Then, the
complexity measures $C^{\alpha \beta }(q)$ vanishes at zero \textit{disorder}
and zero \textit{order, }and has the maximum
\begin{equation}
C_{\max }^{\alpha \beta }=\frac{\alpha ^{\alpha }\beta ^{\beta
}}{(\alpha +\beta )^{(\alpha +\beta )}}\: {\rm at }\:\Delta
_{m}=\frac{\alpha }{\alpha +\beta }, \:\Omega _{m}=\frac{\beta
}{\alpha +\beta }  \label{5}
\end{equation}

\section{\textit{\ }\textbf{Entropies for two-body scattering }}

Therefore, in order to introduce a measure of complexity of the
hadronic system of quantum states produced in hadronic scatterings
we must define the \textit{disorder (}$\Delta $), or
\textit{order} ($\Omega $) for the hadron-hadron scattering. These
physical quantities can be defined just as in Eqs (1)-(2) by using
the scattering Boltzman-Gibs-like entropy [2] or more general
Tsallis-like scattering entropies introduced by us in Refs.
[1,2]. For completeness we start our discussion with $f_{++}(x)$ and $%
f_{+-}(x)$, $x\in [-1,1]$, as the scattering helicity amplitudes of \textit{%
meson-nucleon scattering}

\begin{equation}
M\:(0^{-})+N\:(\frac{1}{2}^{+})\rightarrow M\:(0^{-})+N%
\:(\frac{1}{2}^{+})  \label{6}
\end{equation}
where $x=\cos (\theta ),\theta $ being the c.m. scattering angle. The
normalization of the helicity amplitudes $f_{++}(x)$ and $f_{+-}(x)$ is
chosen such that the c.m. differential cross section $\frac{d\sigma }{%
d\Omega }(x)$ is given by

\begin{equation}
\frac{d\sigma }{d\Omega }(x)=\mid f_{++}(x)\mid ^{2}+\mid f_{+-}(x)\mid ^{2}
\label{7}
\end{equation}

Since we will work at fixed energy, the dependence of $\sigma _{el}$ and,$%
\frac{d\sigma }{d\Omega }(x)$ and of $f(x)$, on this variable was
suppressed. Hence, the helicities of incoming and outgoing nucleons are
denoted by $\mu $, $\mu ^{^{\prime }}$, and was written as (+),(-),
corresponding to $(\frac{1}{2})$ and $(-\frac{1}{2})$, respectively. In
terms of the partial waves amplitudes $f_{J+}$ and $f_{J-}$ we have
\begin{equation}
\begin{tabular}{l}
$f_{++}(x)=\sum_{J=\frac{1}{2}}^{J_{\max }}(J+1/2)(f_{J-}+f_{J+})d_{\frac{1}{%
2}\frac{1}{2}}^{J}(x)$ \\
$f_{+-}(x)=\sum_{J=\frac{1}{2}}^{J_{\max }}(J+1/2)(f_{J-}-f_{J+})d_{-\frac{1%
}{2}\frac{1}{2}}^{J}(x)$%
\end{tabular}
\label{8}
\end{equation}
where the d$_{\mu \nu }^{J}(x)$-rotation functions are given by

\begin{equation}
d_{\frac{1}{2}\frac{1}{2}}^{J}(x)=\frac{1}{l+1}\cdot \left[ \frac{1+x}{2}%
\right] ^{\frac{1}{2}}\left[ \stackrel{\circ }{P}_{l+1}(x)-\stackrel{\circ }{%
P}_{l}(x)\right]  \label{9}
\end{equation}

\begin{equation}
d_{-\frac{1}{2}\frac{1}{2}}^{J}(x)=\frac{1}{l+1}\cdot \left[ \frac{1-x}{2}%
\right] ^{\frac{1}{2}}\left[ \stackrel{\circ }{P}_{l+1}(x)+\stackrel{\circ }{%
P}_{l}(x)\right]  \label{10}
\end{equation}
$P_{l}(x)$ are Legendre Polynomials and prime indicates differentiation with
respect to x $\equiv \cos \theta $.

Now, the elastic integrated cross section for the meson-nucleon scattering
is expressed in terms of partial amplitudes $f_{J+}$ and $f_{J-}$
\begin{equation}
\sigma _{el}/2\pi =\int_{-1}^{+1}dx\frac{d\sigma }{d\Omega }(x)=\sum_{J=%
\frac{1}{2}}^{J_{\max }}(2J+1)(\mid f_{J+}\mid ^{2}+\mid f_{J-}\mid ^{2})
\label{(11)}
\end{equation}

\textbf{J-nonextensive entropy for the quantum scattering}

Now, we define two kind of Tsallis\textit{-}like scattering entropies. One
of them, namely $S_{J}(p),$ p$\in R,$ is special dedicated to the
investigation of the nonextensive statistical behavior of the \textit{%
angular momentum} $J-$\textit{quantum states}, and can be defined by

\begin{equation}
S_{J}(p)=\left[ 1-\sum (2l+1)p_{l}^{p}\right] /(p-1),\smallskip \ \:%
p\in R,\  \label{12}
\end{equation}
where the probability distributions $p_{J},$ is given by

\begin{equation}
p_{J}=\frac{\mid f_{J+}\mid ^{2}+\mid f_{J-}\mid ^{2}}{\sum_{J=\frac{1}{2}%
}^{J_{\max }}(2J+1)(\mid f_{J+}\mid ^{2}+\mid f_{J-}\mid
^{2})},\smallskip \ \sum_{J=\frac{1}{2}}^{J_{\max
}}(2J+1)\:p_{J}=1  \label{13}
\end{equation}

$\theta $\textbf{-nonextensive entropy for the quantum scattering}

In similar way, for the $\theta -$scattering states considered as
statistical canonical ensemble, we can investigate their nonextensive
statistical behavior by using an angular Tsallis-like scattering entropy $%
S_{\theta }(q)$ defined as

\begin{equation}
S_{\theta }(q)=\left[ 1-\int_{-1}^{+1}dx[P(x)]^{q}\right]
/(q-1),\smallskip \ q\in R  \label{(14)}
\end{equation}
where
\begin{equation}
P(x)=\frac{2\pi }{\sigma _{el}}\,\cdot \frac{d\sigma }{d\Omega }(x),\:%
\int_{-1}^{1}P(x)dx=1  \label{15}
\end{equation}
with $\frac{d\sigma }{d\Omega }(x)$ and $\sigma _{el}$ defined by
Eqs.(7)-(11).

The above Tsallis-like scattering entropies posses two important properties.
First, in the limit $k\rightarrow 1,k\equiv p,q,$ the Boltzman-Gibs kind of
entropies is recovered:
\begin{equation}
\begin{tabular}{l}
$\lim_{p\rightarrow 1}S_{J}(p)=S_{J}(1)=-\sum (2J+1)p_{J}\ln p_{J}%
\:$ \\
$\lim_{q\rightarrow 1}S_{\theta }(q)=S_{\theta
}(1)=-\int_{-1}^{+1}dxP(x)\ln P(x)$%
\end{tabular}
\label{16}
\end{equation}

Secondly, these entropies are \textit{nonextensive} in the sense that
\begin{equation}
S_{A+B}(k)=S_{A}(k)+S_{B}(k)+(1-k)S_{A}(k)S_{B}(k),\smallskip \ k=p,q
\label{17}
\end{equation}
for any independent sub-systems $A,B$ $(p_{A+B}=p_{A}\cdot p_{B}).$ Hence,
each of the indices $p\neq 1$ or $q\neq 1$ from the definitions (12) and
(17) can be interpreted as measuring the degree \textit{nonextensivity of
the quantum state system. }

We next consider the \textit{maximum-entropy} (MaxEnt) problem
\begin{equation}
\max \{S_{J}(p),S_{\theta }(q)\}{\rm \: when \:}\sigma _{el}={\rm fixed\: and\: }%
\frac{d\sigma }{d\Omega }(1)={\rm fixed}  \label{18}
\end{equation}
as criterion for the determination of the \textit{equilibrium distributions}
$p_{l}^{me}$ and $P^{me}(x)$ for the quantum states from the $(0^{-}\frac{1}{%
2}\rightarrow 0^{-}\frac{1}{2})-$scattering. For the\textit{\ J-quantum
states, }in\textit{\ }the spin $(0^{-}\frac{1}{2}\rightarrow 0^{-}\frac{1}{2}%
)$ scattering case, these distributions are given by:
\begin{equation}
\begin{tabular}{l}
$p_{J}^{me}=p_{J}^{o1}=\frac{1}{2K_{\frac{1}{2}\frac{1}{2}}(1,1)}=\frac{1}{%
(J_{o}+1)^{2}-1/4},\smallskip \ {\rm \: for }\frac{1}{2}\leq J\leq J_{o},%
{\rm \: and}$ \\
$p_{J}^{me}=0,\smallskip \ {\rm \: for \:}J\geq J_{o}+1$%
\end{tabular}
\label{19}
\end{equation}
while, for the $\theta -$\textit{quantum states}, these distributions are as
follows

\begin{equation}
P^{me}(x)=P^{o1}(x)=\left[ \frac{K_{\frac{1}{2}\frac{1}{2}}^{2}(x,1)}{K_{%
\frac{1}{2}\frac{1}{2}}(1,1)}\right]  \label{20}
\end{equation}
Consequently, by a straightforward calculus we obtain that the solution of
the problem (18) is given by

\begin{equation}
S_{J}^{\max }(p)=S_{J}^{o1}(p)=\left[ 1-[2K_{\frac{1}{2}\frac{1}{2}%
}(1,1)]^{1-p}\right] /(p-1),{\rm for \:}p>0  \label{21}
\end{equation}

\begin{equation}
S_{\theta }^{\max }(q)=S_{\theta }^{o1}(q)=\left[ 1-\int_{-1}^{+1}dx\left(
\frac{\left[ K_{\frac{1}{2}\frac{1}{2}}(x,1)\right] ^{2}}{K_{\frac{1}{2}%
\frac{1}{2}}(1,1)}\right) ^{q}\right] /(q-1),{\rm \: for \:}q>0
\label{22}
\end{equation}
where the\textit{\ reproducing kernel function } $K_{\frac{1}{2}\frac{1}{2}%
}(x,1)$ is given by
\begin{equation}
K_{\frac{1}{2}\frac{1}{2}}(x,1)=\frac{1}{2}\sum_{1/2}^{J_{o}}(2J+1)d_{\frac{1%
}{2}\frac{1}{2}}^{J}(x)  \label{23}
\end{equation}
while the \textit{optimal angular momentum} $J_{o\:}$ is
\begin{equation}
(J_{o}+1)^{2}-1/4=2K_{\frac{1}{2}\frac{1}{2}}(1,1)=\frac{4\pi }{\sigma _{el}}%
\frac{d\sigma }{d\Omega }(1)  \label{24}
\end{equation}
$d_{\frac{1}{2}\frac{1}{2}}^{J}(x)$ are the usual spin 1/2-rotation
functions.

\textit{Proof:} The proof of the problem (18) is equivalent to the following
unconstrained maximization problem:
\begin{equation}
\pounds \rightarrow \max  \label{25}
\end{equation}
where the Lagrangian function is defined as

\begin{eqnarray}
\pounds &\equiv &\lambda _{0}\left\{ S_{J}(p),S_{\theta }(q)\right\}
+\lambda _{1}\left\{ \sigma _{el}/4\pi -\sum (2J+1)\left[ \mid f_{J-}\mid
^{2}+\mid f_{J+}\mid ^{2}\right] \right\} +  \label{26} \\
&&\lambda _{2}\left\{ \frac{d\sigma }{d\Omega }(1)-\left[ \sum (2J+1){\rm Re}%
f_{J+}\right] ^{2}-\left[ \sum (2J+1){\rm Im}f_{J+}\right]
^{2}\right\} \nonumber
\end{eqnarray}

Hence, the solution of the problem (18), correspond to the the singular case
$\lambda _{0}=0,$ and is reduced just to the solution of the \textit{minimum
constrained distance in space of quantum states:}
\begin{equation}
\min \sum (2J+1)\left[ \mid f_{J-}\mid ^{2}+\mid f_{J+}\mid
^{2}\right] {\rm \: when \:}\frac{d\sigma }{d\Omega }(1)={\rm
\:is\: \: fixed} \label{27}
\end{equation}
from which we obtain the following analytic form of the optimal scattering
helicity amplitudes

\begin{equation}
f_{++}^{o1}(x)=f_{++}(1)\cdot \frac{K_{\frac{1}{2}\frac{1}{2}}(x,1)}{K_{%
\frac{1}{2}\frac{1}{2}}(1,1)},\:f_{+-}^{o1}(x)=0,  \label{28}
\end{equation}
where the \textit{reproducing kernels}
$K_{\frac{1}{2}\frac{1}{2}}(x,1)$ are given by Eqs. (23)-(24). All
these results together other are presented in Table 1

\section{\textbf{Complexity measure for quantum scattering }\textit{\ }}

Now, let us apply the definitions (1) and (4) of the\textit{\ disorder,
order, }and\textit{\ complexity measure} to the nonextensive quantum
scattering of the elementary particles. Then, we define the scattering ($%
J,\theta $)$-$\textit{disorders} \textit{\ }as follows

\begin{equation}
\Delta _{X}(q)=\frac{S_{X}(q)}{S_{X}^{\max }(q)},\smallskip \ \smallskip \
X\equiv \theta ,J  \label{29}
\end{equation}
and the corresponding scattering ($\theta ,J$)$-$\textit{orders }by the
relations
\begin{equation}
\Omega _{X}(q)=1-\Delta _{X}(q),\smallskip \ \smallskip \ X\equiv \theta ,J
\label{30}
\end{equation}
where $S_{\theta }(q),S_{J}(q)$ are the scattering \textit{Tsallis-like
scattering entropies} defined in Section 3.

Therefore, using Eqs. (29)-(30) and (12)-(15) we can calculate the\textit{\
scattering complexities} $C_{X}^{\alpha \beta }(q)$ defined as follows
\begin{equation}
C_{X}^{\alpha \beta }(q)=\Delta _{X}^{\alpha }(q)\Omega
_{X}^{\beta }(q),\smallskip \ X\equiv \theta ,J{\: and \:}q\in R
\label{31}
\end{equation}

\section{\textbf{Numerical Results}}

Now, for a systematic experimental investigation of the saturation of the
\textit{optimality limits in hadron-hadron scattering }is necessary to use
the formulas from the Table 1 and the available experimental phase-shifts
[8-10] to solve the following numerical problems: (i) To reconstruct the
experimental pion-nucleon, kaon-nucleon and antikaon-nucleon scattering
amplitudes; (ii) To obtain numerical values of the experimental scattering
entropies $S_{J}(q),$ $S_{\theta }(q),\overline{S}_{J\theta }(p,q)$ from the
reconstructed amplitudes; (iii) To obtain the numerical values of the
\textit{optimal angular momentum J}$_{o}$ from experimental scattering
amplitudes and to calculate the numerical values for the PMD-SQS-optimal
entropies $S_{J}^{o1}(q),$ $S_{\theta }^{o1}(q);\overline{S}_{J\theta }(p,q);
$ (iv) To obtain numerical values for $\chi _{J}^{2}(p)$ or/and $\chi
_{\theta }^{2}(q)$-\textit{test functions} given by
\begin{equation}
\chi _{X}^{2}(k)=\sum_{i=1}^{n_{\exp }}\left[ \frac{%
[S_{X}(k)]_{i}-[S_{X}^{o1}(k)]_{i}}{[\Delta S_{X}^{o1}]_{i}}\right]
^{2},\quad X\equiv J,\theta ;\stackrel{-}{J\theta }\;\;k\equiv p,q
\label{32}
\end{equation}
where
\begin{equation}
\Delta S_{X}^{o1}(k)=|[S_{X}^{o1}(k)]_{J_{o}+1}-[S_{X}^{o1}(k)]_{J_{o}-1}|
\label{33}
\end{equation}
are the values of the $PMD-SQS-$\textit{optimal entropies} $%
[S_{X}^{o1}(k)]_{J_{o}\pm 1}$ calculated with the optimal angular momenta $%
J_{o}\pm 1,$ respectively. This procedure is equivalent with
assumption of an error of $\Delta J_{o}=\pm 1$ in estimation of
the experimental values of the optimal angular momentum $J_{o}.$
The results obtained in this way are presented in the Figs. 1 and
Tables 2-4. So, the best fit is obtained (see Tables 2) for the
correlated pairs $p$ and $q=p/(2p-1)$ with the values of $p $ in
the range $p=$ $0.6$ and $q=p/(2p-1)=3$.0. However, at this level
of error analysis the case of the extensive system (p=q=1) of
quantum scattering states is also consistent with the experimental
data while the pair of nonextensivities (p=3.0,q=0.6) rejected by
the values of $\chi ^{2}/n_{D}-$test function (see Table 4).

Moreover, in Fig. 2-3 we presented the numerical results for the \textit{%
complexity of the J-quantum systems of states} produced in the pion-nucleon
scattering. So, the experimental values of the complexities $C_{J}^{11}(p)$
for $\pi ^{+}p$ and $\pi ^{-}p-$scatterings, obtained from the available
experimental phase-shifts [19], are compared with the \textit{%
PMD-SQS-optimal state} predictions (full curve
[$C_{J}^{11}(p)]^{o1}=0$). In all cases, the error-bands (the grey
regions) around the optimality curves are calculated by assuming
an error of $\Delta J_{o}=\pm 1$ in calculation of the optimal
angular momentum J$_{o}$ From the results of the Figs.2-3 we
conclude that the scattering complexities $C_{J}^{11}(p)$ are in
excellent agreement with the optimal complexities
[$C_{J}^{11}(p)]^{o1}=0$. So, a  \textit{principle of minimum
complexity in hadronic interaction} is suggested with high
accuracy ($CL>99\%$) by the experimental pion-nucleon scattering
data.

\section{Summary and Conclusions }

In this paper, by using \textit{the optimality  [}1-4\textit{] and }$%
[S_{J}(p)$, $S_{\theta }(q)]$-Tsallis-like entropies ,  the  \textit{%
complexity}  and nonextensivity of the [$J$ and $\theta $]-quantum states
produced in hadronic scatterings are  investigated.  The main results and
conclusions can be summarized as follows:

\begin{itemize}
\item[(i)]  Using the available experimental phase shifts analysis we
calculated the numerical values for the [$S_{J}(p)$, $S_{\theta }(q))$]-%
\textit{Tsallis-like scattering entropies }for the pure isospin $I-$%
scattering states: [$(\pi N)_{I=1/2,3/2}$; $(KN)_{I=0,1}$; $(\overline{K}%
N)_{I=0,1}$];

\item[(ii)]  We presented strong experimental evidence for the \textit{%
saturation of the} $[S_{J}^{o1}(p),S_{\theta
}^{o1}(q)]-$\textit{PMD-SQS optimal limits} for all nonextensive
($J$, $\theta $)-statistical ensembles of quantum states produced
in hadron-hadron scattering (see Fig.1 and Table
2). These results allow to conclude that the $[J]$-\textit{quantum} \textit{%
system} and $[\theta ]$-\textit{quantum} \textit{system} are
produced at "\textit{equilibrium}" but with the
$[\frac{1}{2p}+\frac{1}{2q}]-$ conjugated nonextensivities $p=0.6$
and $q=p/(2p-1)=3$ in all investigated
isospin scattering states: [$(\pi N)_{I=1/2,3/2};(KN)_{I=0,1};(\overline{K}%
N)_{I=0,1}]$. So the ''geometric origin'' of the nonextensivities
p and q (as dimensions of the Hilbert spaces $L_{2p}$ and
$L_{2q})$ as well as their correlations are experimentally
confirmed with high accuracy ($CL>99\%$). The strong experimental
evidence obtained here for the \textit{nonextensive statistical
behavior} of the $(J,\theta )-$ quantum scatterings states in the
pion-nucleon, kaon-nucleon and antikaon-nucleon scatterings can be
interpreted as an indirect manifestation of \textit{the presence
of the quarks and gluons } as fundamental constituents of the
scattering system having the strong-coupling long-range regime
required by the Quantum Chromodynamics.

\item[(iii)]  We presented strong experimental evidence fo the
validity of \textit{a principle of minimum complexity (PMC) in
pion-nucleon scattering. PMC} is supported with high accuracy
($CL>99\%$) by all the experimental data especially at energies
higher than 2 GeV beyond the resonance region.
\end{itemize}

Finally, we note that further investigations are needed since this \textit{%
saturation of optimality and complexity limits} as well as \textit{%
nonextensive statistical behavior} of the quantum scattering, evidentiated
here with high accuracy $(CL>99\%)$, can be a signature for a \textit{new
universal law }of the quantum scattering.

\newpage

\begin{center}
\textbf{TABLE\ CAPTIONS}
\end{center}

\begin{itemize}
\item[\textbf{Table 1:}]  The optimal distributions, reproducing kernels,
optimal entropies, entropic bands, for the $(0^{-}1/2^{+}\rightarrow
0^{-}1/2^{+})-$scatterings.

\item[\textbf{Table 2:}]  $\chi ^{2}/n_{D}$ obtained from comparisons of the
experimental scattering entropies: S$_{J}(p),$ S$_{\theta }(q)$ with the
optimal entropies: S$_{J}^{o1}(p),$ S$_{\theta }^{o1}(q),$respectively, for
the pair (p=0.6, q=3.0).

\item[\textbf{Table 3:}]  $\chi ^{2}/n_{D}$ obtained from comparisons of the
experimental scattering entropies: S$_{J}(p),$ S$_{\theta }(q)$ with the
optimal entropies: S$_{J}^{o1}(p),$ S$_{\theta }^{o1}(q),$ respectively, for
the pair (p=1.0, q=1.0).

\item[\textbf{Table 4:}]  $\chi ^{2}/n_{D}$ obtained from comparisons of the
experimental scattering entropies: S$_{J}(p),$ S$_{\theta }(q)$ with the
optimal entropies: S$_{J}^{o1}(p),$ S$_{\theta }^{o1}(q),$ respectively, for
the pair (p=3.0, q=0.6).
\end{itemize}

\begin{center}
\textbf{FIGURES\ CAPTIONS}
\end{center}

\begin{itemize}
\item[\textbf{Fig. 1:}]  The experimental values of the Tsallis-like
scattering entropies $S_{J}(p)$ (12) for $[(\pi N)_{I=1/2,3/2}$; $%
(KN)_{I=0,1}$; $(\overline{K}N)_{I=0,1}]-$ scatterings, obtained from the
available experimental phase-shifts [8-10], are compared with the
PMD-SQS-optimal state predictions $S_{J}^{o1}(p)$ (full curve).The
error-band (the grey region) around the optimality curve is calculated by
assuming an error of $\Delta J_{o}=\pm 1$ in calculation of the optimal
angular momentum J$_{o}$.

\item[\textbf{Fig. 2:}]  The experimental values of the complexities $%
C_{J}^{11}(p)$ for $\pi ^{+}p-$scatterings, obtained, by using
Eqs. (29)-(31), from the available experimental phase-shifts
[8-10], are compared with the PMD-SQS-optimal state complexity
(full curve $C_{J}^{11}(p)=0).$ The error-band (the grey region)
around the optimality curve is calculated by assuming an error of
$\Delta J=\pm 1$ in calculation of the optimal angular momentum
J$_{o}$.

\item[\textbf{Fig. 3:}]  The experimental values of the complexities $%
C_{J}^{11}(p)$ for $\pi ^{-}p-$scatterings, obtained from the available
experimental phase-shifts [8], are compared with the PMD-SQS-optimal state
complexity (the full curve $C_{J}^{11}(p)=0).$ The error-band (the grey
region) around the optimality curve is calculated by assuming an error of $%
\Delta J=\pm 1$ in calculation of the optimal angular momentum J$_{o}$.
\end{itemize}

\begin{center}
\textbf{Table 1}

\begin{tabular}[t]{|c|c|c|}
\hline
Nr. & Name & $(0^{-}1/2^{+}\rightarrow 0^{-}1/2^{+})-$scatterings \\ \hline
1 & Optimal inequalities & $\frac{d\sigma }{d\Omega }(x)\leq K_{\frac{1}{2}%
\pm \frac{1}{2}}(\pm 1,\pm 1)\parallel f\parallel ^{2}$ \\ \hline
2 & Optimal states & $f^{o+1}(x)=f(1)\frac{K_{\frac{1}{2}\frac{1}{2}}(x,+1)}{%
K_{\frac{1}{2}\frac{1}{2}}(1,1)},$\smallskip \ $f^{o+1}(x)=f(-1)\frac{K_{-%
\frac{1}{2}\frac{1}{2}}(x,-1)}{K_{-\frac{1}{2}\frac{1}{2}}(-1,-1)}$ \\ \hline
2 & $
\begin{array}{c}
\mathrm{Reproducing\;kernels} \\
K_{\frac{1}{2}\pm \frac{1}{2}}(x,\pm 1)
\end{array}
$ & $
\begin{tabular}{c}
$K_{\frac{1}{2}\frac{1}{2}}(x,+1)=\sum_{\frac{1}{2}}^{J_{o}}(J+\frac{1}{2}%
)d_{\frac{1}{2}\frac{1}{2}}^{J}(x)$ \\
$K_{\frac{1}{2}-\frac{1}{2}}(x,-1)=\sum_{\frac{1}{2}}^{J_{o}}(J+\frac{1}{2}%
)d_{\frac{1}{2}-\frac{1}{2}}^{J}(x)$ \\
$2K_{\frac{1}{2}\pm \frac{1}{2}}(\pm 1,\pm 1)=(J_{o}+1)^{2}-1/4$%
\end{tabular}
$ \\ \hline
3 & $
\begin{array}{c}
\mathrm{{Optimal\;distributions\mathit{\ }}} \\
P^{o\pm 1}(x)
\end{array}
$ & $P^{o1}(x)=\frac{\left[ K_{\frac{1}{2}\frac{1}{2}}(x,1)\right] ^{2}}{K_{%
\frac{1}{2}\frac{1}{2}}(1,1)},\smallskip \ P^{o-1}(x)=\frac{\left[ K_{\frac{1%
}{2}-\frac{1}{2}}(x,-1)\right] ^{2}}{K_{\frac{1}{2}-\frac{1}{2}}(-1,-1)}$ \\
\hline
4 & $
\begin{array}{c}
\mathrm{Optimal\;distribution} \\
\{p_{J}^{o1}\}
\end{array}
$ &
\begin{tabular}{c}
$p_{J}^{o\pm 1}=\frac{1}{2K_{\frac{1}{2}\pm \frac{1}{2}}(\pm 1,\pm 1)},$ for
$1/2\leq J\leq J_{o}$ \\
$p_{J}^{o\pm 1}=0$ for $J\geq J_{o}+1$%
\end{tabular}
\\ \hline
5 & $
\begin{array}{c}
\mathrm{Number\;of} \\
\mathrm{optimal\;states} \\
\mathrm{\;for\;y=1}
\end{array}
$ & $N_{o}=\sum (2J+1)=(J_{o}+1)^{2}-1/4=2K_{\frac{1}{2}\pm \frac{1}{2}}(\pm
1,\pm 1)$ \\ \hline
6 & $
\begin{array}{c}
\mathrm{Optimal} \\
\mathrm{angular-momentum}
\end{array}
$ & $J_{o}=\left\{ \left[ \frac{4\pi }{\sigma _{el}}\frac{d\sigma }{d\Omega }%
(\pm 1)+\frac{1}{4}\right] ^{1/2}-1\right\} $ \\ \hline
7 & $
\begin{array}{c}
\mathrm{Optimal\;entropy} \\
S_{J}^{o1}(q)
\end{array}
$ & $S_{J}^{o\pm 1}(p)=\frac{1}{p-1}\left[ 1-N_{os}^{1-p}\right] $ \\ \hline
8 & $
\begin{array}{c}
\mathrm{Optimal\;entropy} \\
S_{\theta }^{o1}(q)
\end{array}
$ & $S_{\theta }^{o\pm 1}(q)=\frac{1}{q-1}\left[ 1-\int_{-1}^{+1}dx\left(
\frac{\left[ K_{\frac{1}{2}\pm \frac{1}{2}}(x,\pm 1)\right] ^{2}}{K(\pm
1,\pm 1)}\right) ^{q}\right] $ \\ \hline
11 & $J-$entropic band & $0\leq S_{J}(p)\leq S_{J}^{o\pm 1}(p)$ \\ \hline
12 & $\theta -$entropic band & $\frac{1}{q-1}[1-K(1,1)^{q-1}]\leq S_{\theta
}(q)\leq S_{\theta }^{o\pm 1}(q)$ \\ \hline
\end{tabular}

\bigskip

\textbf{Table 2}

\begin{tabular}{||c||c||c||c||c||}
\hline\hline
&
\begin{tabular}{c}
Hadron-hadron \\
scattering
\end{tabular}
&
\begin{tabular}{c}
Number \\
of PSA
\end{tabular}
&
\begin{tabular}{cc}
p & q=$\frac{p}{2p-1}$%
\end{tabular}
&
\begin{tabular}{c}
$\chi ^{2}/n_{D}$ \\
\begin{tabular}{cc}
S$_{J}(p)$ & S$_{\theta }(q)$%
\end{tabular}
\end{tabular}
\\ \hline\hline
1 & $\pi N\rightarrow (\pi N)_{I=1/2}$ & 88 &
\begin{tabular}{c}
0.6
\end{tabular}
\begin{tabular}{c}
3.0
\end{tabular}
&
\begin{tabular}{cc}
0.102 & 0.015
\end{tabular}
\\ \hline\hline
2 & $\pi N\rightarrow (\pi N)_{I=3/2}$ & 88 &
\begin{tabular}{c}
0.6
\end{tabular}
\begin{tabular}{c}
3.0
\end{tabular}
&
\begin{tabular}{cc}
0.130 & 0.090
\end{tabular}
\\ \hline\hline
3 & $KN\rightarrow (KN)_{I=0}$ & 52 &
\begin{tabular}{c}
0.6
\end{tabular}
\begin{tabular}{c}
3.0
\end{tabular}
&
\begin{tabular}{cc}
0.449 & 0.035
\end{tabular}
\\ \hline\hline
4 & $KN\rightarrow (KN)_{I=1}$ & 53 &
\begin{tabular}{c}
0.6
\end{tabular}
\begin{tabular}{c}
3.0
\end{tabular}
&
\begin{tabular}{cc}
0.089 & 0.045
\end{tabular}
\\ \hline\hline
5 & $\overline{K}N\rightarrow (\overline{K}N)_{I=0}$ & 50 &
\begin{tabular}{c}
0.6
\end{tabular}
\begin{tabular}{c}
3.0
\end{tabular}
&
\begin{tabular}{cc}
0.168 & 0.009
\end{tabular}
\\ \hline\hline
6 & $\overline{K}N\rightarrow (\overline{K}N)_{I=1}$ & 50 &
\begin{tabular}{c}
0.6
\end{tabular}
\begin{tabular}{c}
3.0
\end{tabular}
&
\begin{tabular}{cc}
0.062 & 0.010
\end{tabular}
\\ \hline\hline
\end{tabular}

\textbf{Table 3}

\begin{tabular}{||c||c||c||c||}
\hline\hline
\begin{tabular}{c}
Hadron-hadron \\
scattering
\end{tabular}
&
\begin{tabular}{c}
Number \\
of PSA
\end{tabular}
&
\begin{tabular}{cc}
p & q=$\frac{p}{2p-1}$%
\end{tabular}
&
\begin{tabular}{c}
$\chi ^{2}/n_{D}$ \\
\begin{tabular}{cc}
S$_{J}(p)$ & S$_{\theta }(q)$%
\end{tabular}
\end{tabular}
\\ \hline\hline
$
\begin{tabular}{c}
$\pi N\rightarrow (\pi N)_{I=1/2}$%
\end{tabular}
$ & 88 &
\begin{tabular}{ll}
1 & 1
\end{tabular}
&
\begin{tabular}{cc}
0.649 & 0.648
\end{tabular}
\\ \hline\hline
\begin{tabular}{c}
$\pi N\rightarrow (\pi N)_{I=3/2}$%
\end{tabular}
& 88 &
\begin{tabular}{ll}
1 & 1
\end{tabular}
&
\begin{tabular}{cc}
0.691 & 1.059
\end{tabular}
\\ \hline\hline
\begin{tabular}{c}
$KN\rightarrow (KN)_{I=0}$%
\end{tabular}
& 52 &
\begin{tabular}{ll}
1 & 1
\end{tabular}
&
\begin{tabular}{cc}
0.146 & 0.494
\end{tabular}
\\ \hline\hline
\begin{tabular}{c}
$KN\rightarrow (KN)_{I=1}$%
\end{tabular}
& 53 &
\begin{tabular}{ll}
1 & 1
\end{tabular}
&
\begin{tabular}{cc}
0.259 & 0.586
\end{tabular}
\\ \hline\hline
\begin{tabular}{c}
$\overline{K}N\rightarrow (\overline{K}N)_{I=0}$%
\end{tabular}
& 50 &
\begin{tabular}{ll}
1 & 1
\end{tabular}
&
\begin{tabular}{cc}
0.199 & 0.267
\end{tabular}
\\ \hline\hline
$\overline{K}N\rightarrow (\overline{K}N)_{I=1}$ & 50 &
\begin{tabular}{ll}
1 & 1
\end{tabular}
&
\begin{tabular}{cc}
0.064 & 0.196
\end{tabular}
\\ \hline\hline
\end{tabular}

\textbf{Table 4}

\begin{tabular}{||c||c||c||c||}
\hline\hline
\begin{tabular}{c}
Hadron-hadron \\
scattering
\end{tabular}
&
\begin{tabular}{c}
Number \\
of PSA
\end{tabular}
&
\begin{tabular}{cc}
p & q=$\frac{p}{2p-1}$%
\end{tabular}
&
\begin{tabular}{c}
$\chi ^{2}/n_{D}$ \\
\begin{tabular}{cc}
S$_{J}(p)$ & S$_{\theta }(q)$%
\end{tabular}
\end{tabular}
\\ \hline\hline
$\pi N\rightarrow (\pi N)_{I=1/2}$ & 88 &
\begin{tabular}{c}
3.0
\end{tabular}
\begin{tabular}{c}
0.6
\end{tabular}
&
\begin{tabular}{cc}
143.6 & 2.181
\end{tabular}
\\ \hline\hline
$\pi N\rightarrow (\pi N)_{I=3/2}$ & 88 &
\begin{tabular}{c}
3.0
\end{tabular}
\begin{tabular}{c}
0.6
\end{tabular}
&
\begin{tabular}{cc}
209.2 & 3.010
\end{tabular}
\\ \hline\hline
$KN\rightarrow (KN)_{I=0}$ & 52 &
\begin{tabular}{c}
3.0
\end{tabular}
\begin{tabular}{c}
0.6
\end{tabular}
&
\begin{tabular}{cc}
0.190 & 1.089
\end{tabular}
\\ \hline\hline
$KN\rightarrow (KN)_{I=1}$ & 53 &
\begin{tabular}{c}
3.0
\end{tabular}
\begin{tabular}{c}
0.6
\end{tabular}
&
\begin{tabular}{cc}
30.53 & 1.030
\end{tabular}
\\ \hline\hline
$\overline{K}N\rightarrow (\overline{K}N)_{I=0}$ & 50 &
\begin{tabular}{c}
3.0
\end{tabular}
\begin{tabular}{c}
0.6
\end{tabular}
&
\begin{tabular}{cc}
11.38 & 0.551
\end{tabular}
\\ \hline\hline
$\overline{K}N\rightarrow (\overline{K}N)_{I=1}$ & 50 &
\begin{tabular}{c}
3.0
\end{tabular}
\begin{tabular}{c}
0.6
\end{tabular}
&
\begin{tabular}{cc}
16.00 & 0.391
\end{tabular}
\\ \hline\hline
\end{tabular}
\end{center}

\begin{figure*}[!]
\includegraphics[height=16cm]{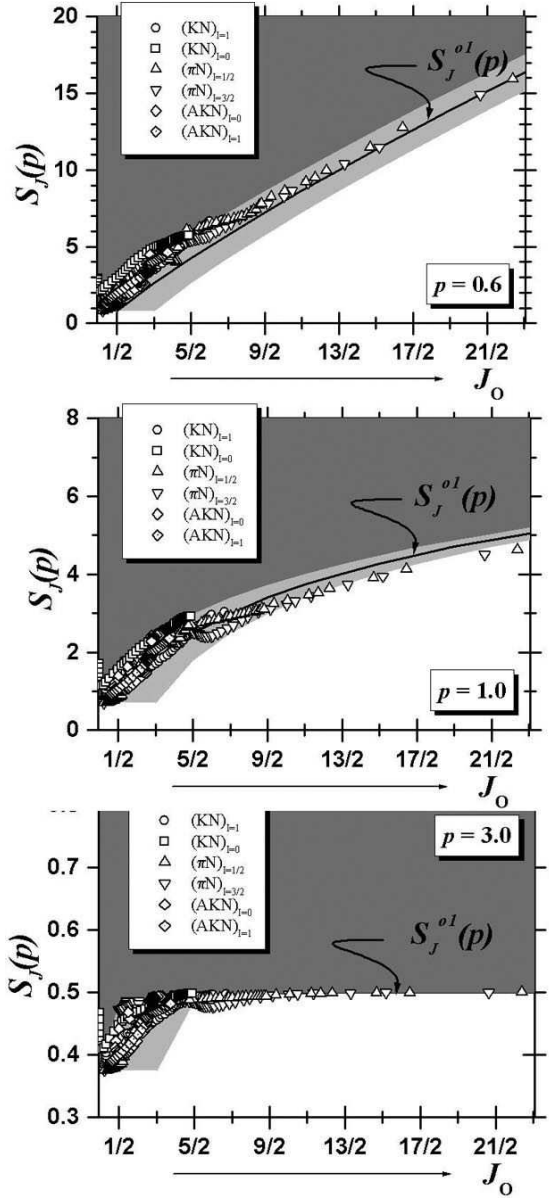}\\
\caption{\label{fig1} \footnotesize The experimental values of the
Tsallis-like
scattering entropies $S_{J}(p)$ (12) for $[(\pi N)_{I=1/2,3/2}$; $%
(KN)_{I=0,1}$; $(\overline{K}N)_{I=0,1}]-$ scatterings, obtained
from the available experimental phase-shifts [8-10], are compared
with the PMD-SQS-optimal state predictions $S_{J}^{o1}(p)$ (full
curve).The error-band (the grey region) around the optimality
curve is calculated by assuming an error of $\Delta J_{o}=\pm 1$
in calculation of the optimal angular momentum J$_{o}$.}
\end{figure*}

\begin{figure*}[!]
\includegraphics[height=16cm]{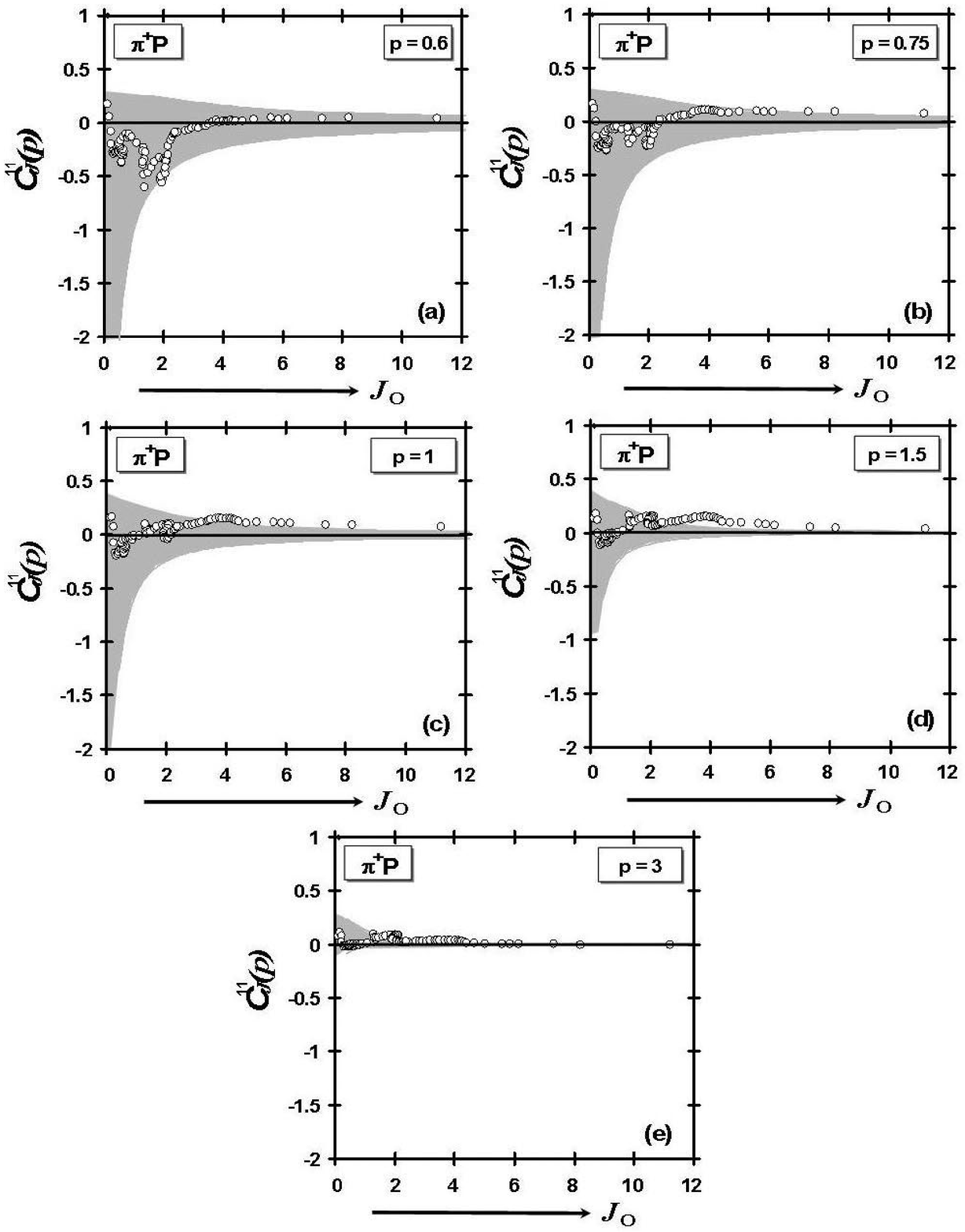}\\
\caption{\label{fig2} \footnotesize The experimental values of the complexities $%
C_{J}^{11}(p)$ for $\pi ^{+}p-$scatterings, obtained, by using
Eqs. (29)-(31), from the available experimental phase-shifts
[8-10], are compared with the PMD-SQS-optimal state complexity
(full curve $C_{J}^{11}(p)=0).$ The error-band (the grey region)
around the optimality curve is calculated by assuming an error of
$\Delta J=\pm 1$ in calculation of the optimal angular momentum
J$_{o}$.}
\end{figure*}
\begin{figure*}[!]
\includegraphics[height=16cm]{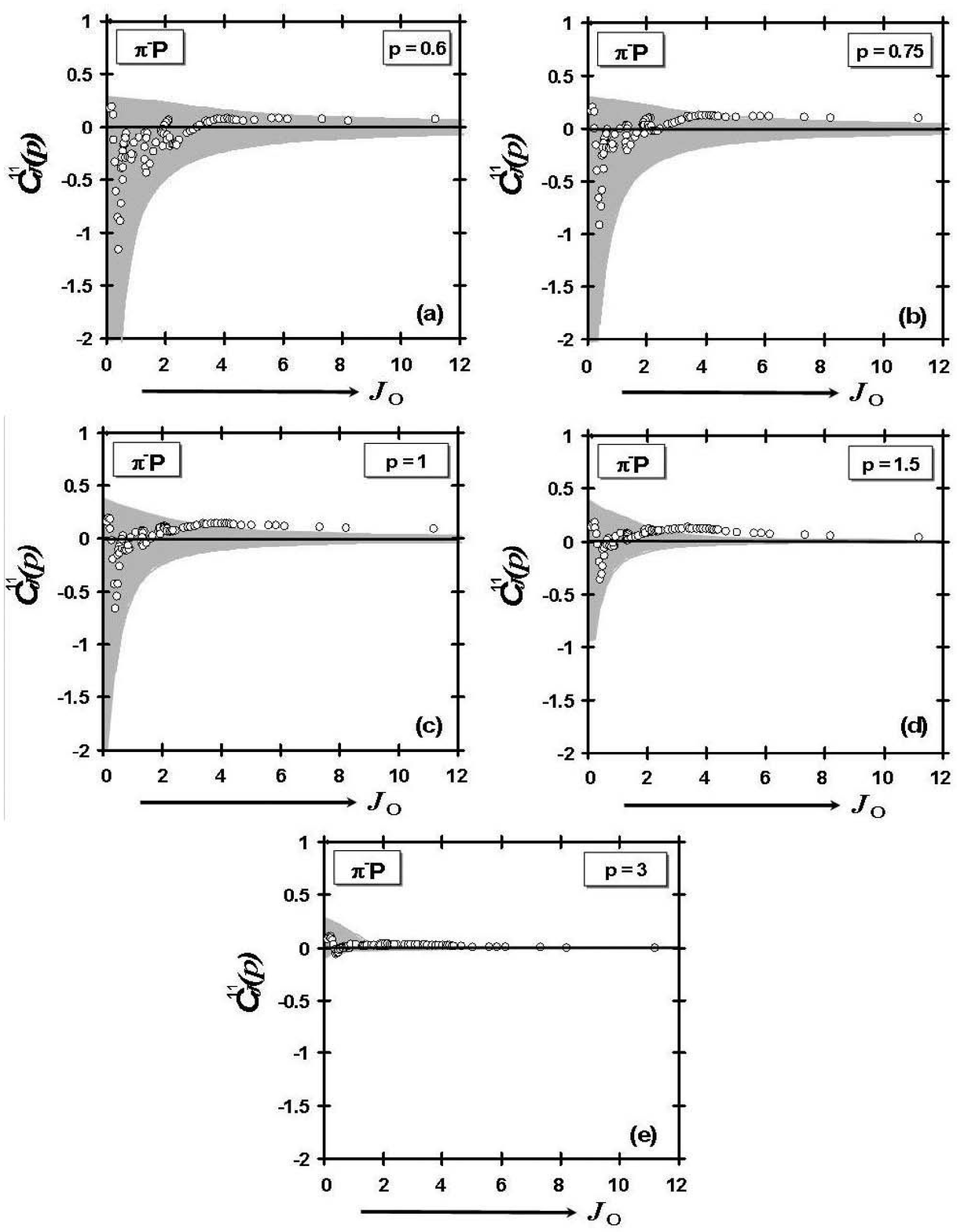}\\
\caption{\label{fig3} \footnotesize The experimental values of the complexities $%
C_{J}^{11}(p)$ for $\pi ^{-}p-$scatterings, obtained from the
available experimental phase-shifts [8], are compared with the
PMD-SQS-optimal state complexity (the full curve
$C_{J}^{11}(p)=0).$ The error-band (the grey
region) around the optimality curve is calculated by assuming an error of $%
\Delta J=\pm 1$ in calculation of the optimal angular momentum
J$_{o}$.}
\end{figure*}
\end{document}